\documentclass[prl,twocolumn,notitlepage]{revtex4-2}
\usepackage[english]{babel}
\usepackage{amsfonts}
\usepackage{physics}
\usepackage{graphicx}
\usepackage{placeins}
\usepackage{url}

\begin{document}
\title{Brownian motion in a growing population of ballistic particles}
\author{Nathaniel V.~Mon P\`ere}
\email[Correspondence to: ]{n.monpere@qmul.ac.uk}

\affiliation{Centre for Cancer Genomics and Computational Biology, Barts Cancer
Institute, Charterhouse Square, London EC1M 6BQ, United Kingdom}
\author{Pierre de Buyl}
\thanks{These authors contributed equally.}
\affiliation{Royal Meteorological Institute of Belgium, Avenue Circulaire 3,
1180 Brussels, Belgium} \affiliation{KU Leuven, Institute for Theoretical
Physics, Celestijnenlaan 200d - box 2415, 3001 Leuven, Belgium}
\author{Sophie de Buyl}
\thanks{These authors contributed equally.}
\affiliation{Applied Physics Research Group, Physics Department, Vrije
Universiteit Brussel, Brussels 1050, Belgium} \affiliation{Interuniversity
Institute of Bioinformatics in Brussels, Vrije Universiteit Brussel --
Universit\'e libre de Bruxelles, Brussels 1050, Belgium}

\begin{abstract}
We investigate the motility of a growing population of cells in a idealized
setting: we consider a system of hard disks in which new particles are added
according to prescribed growth kinetics, thereby dynamically changing the number
density. As a result, the expected Brownian motion of the hard disks is
modified. We compute the density-dependent friction of the hard disks and insert
it in an effective Langevin equation to describe the system, assuming that the
inter-collision time is smaller than the timescale of the growth.
We find that the effective Langevin description captures the changes in motility
in agreement with the simulation results. Our framework can be extended to other
systems in which the transport coefficient varies with time.
\end{abstract}

\maketitle

\section{Introduction}

The statistical description of the movement of living organisms enables their
quantitative study. Notable examples include the study of animal
migration~\cite{pearson_1906} and the motion of
bacteria~\cite{berg_random_walks_biology_1983}.
Over the past decades there has been increasing interest in self-propelled
particles whose ability to provide their own propulsion results in systems where
time-reversibility and energy conservation cannot always be satisfied
\cite{cates_diffusive_2012, romanczuk_active_2012}. Furthermore, including
interactions between organisms or with the environment can lead to the emergence
of biological phase transitions and self-organization \cite{fily_athermal_2012,
tailleur_statistical_2008, stenhammar_continuum_2013, vicsek1995novel}. Besides
sourcing energy for their own motion, another important characteristic of many
such organisms is their ability to reproduce. This can under some circumstances
impact the properties of motion of an active population, as the amount of
interactions taking place will depend on the population density at a given time.
One conspicuous example of such a system is a culture of motile cells on a
spatially confined substrate. This type of in vitro setup is not uncommon,
employed for example in the study of human cell migration
\cite{lintz_mechanics_2017, wu_biophysics_2018} related to tasks such as tissue
growth \cite{weijer_collective_2009}, wound healing
\cite{friedl_collective_2009}, and vascularization \cite{risau_mechanisms_1997},
as well as in studies concerning the migration of tumour cells in metastatic
cancer \cite{lin_role_2019}. In many such cases the motility of the individual
cells is inextricably linked to the surrounding population growth, and
measurements of related statistical quantities should be interpreted in the
context of this dependence.

Here we introduce a simple parameter-free generalizable model for the averaged
effect of growth in a system of repulsively interacting particles.
Based on the notion of particles obstructing one another's paths, we adapt the
Langevin equation for Brownian motion to include a dependence on the particle
volume density. We show how this formalism effectively models the statistics of
a 2D gas of hard-disks subject to particle number growth through comparison with
stochastically seeded simulations of ballistically moving hard disks undergoing
elastic interactions. Because this approach remains agnostic as to both the
characteristics of the particles' motion between interactions as well as the
mechanism by which the particle density varies, it is potentially applicable in
any situation where there is an interest in studying properties of statistical
motility in a population subject to density dynamics.

\section{Modeling and methods}

\subsection{Dynamical model}

\begin{figure}
    \centering
    \includegraphics[width=0.45\textwidth]{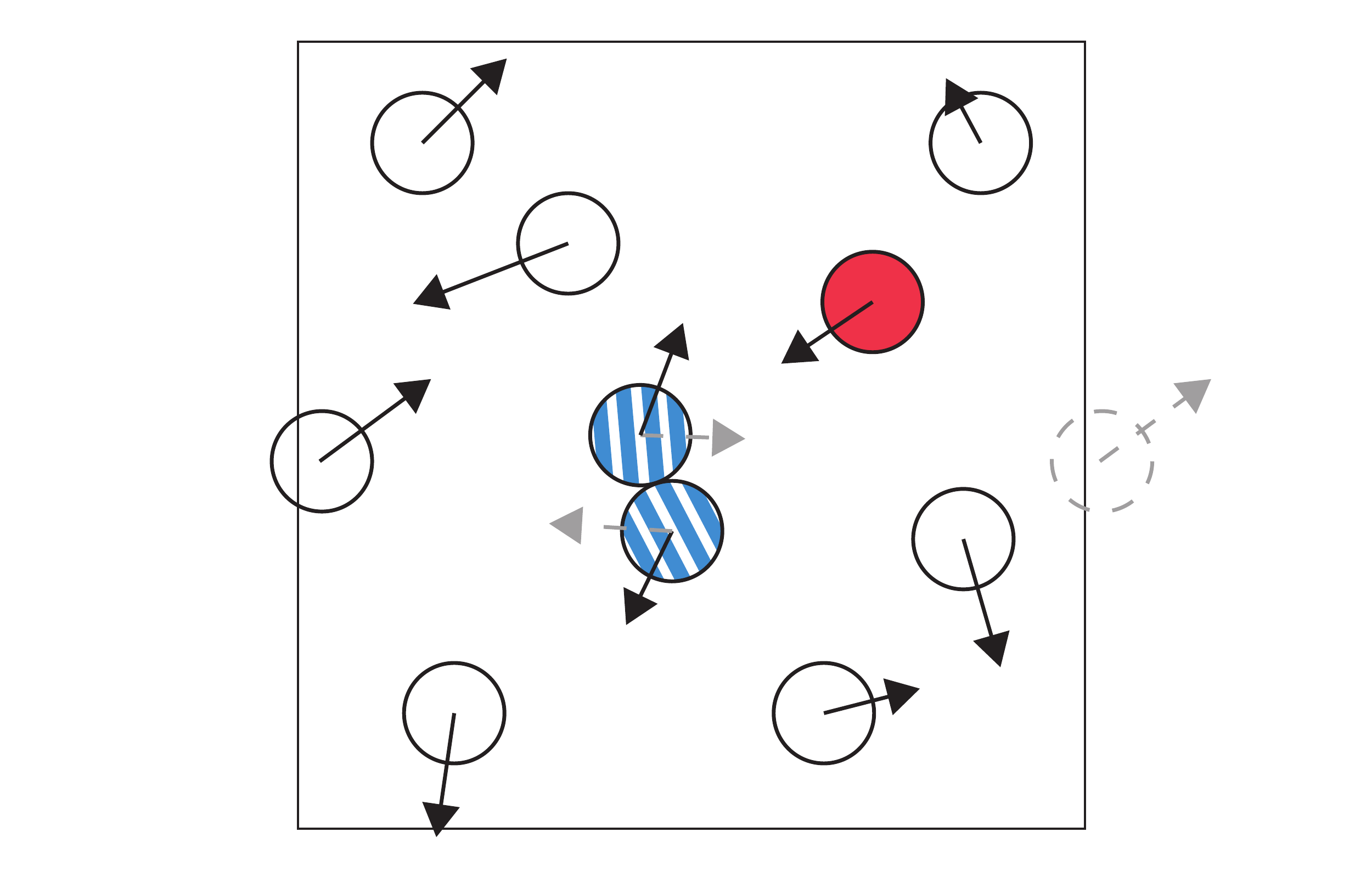}
    \caption{ Schematic illustration of the model. The particles (represented as circles) are hard disks of finite area moving at fixed speed in the periodic domain (the dashed greyed-out particle is re-inserted at its periodic location). Their ballistic trajectories are interrupted by momentum-conserving collisions (blue semi-filled circles). New particles (red filled circle) are inserted at random non-occupied positions, with velocity given by the corresponding thermal distribution, and at times prescribed by the growth kinetics.}
    \label{fig:schematic}
\end{figure}

We consider a two-dimensional system where the Brownian particles are hard disks
with diameter $d$ and mass $m$ each occupying a finite surface $\pi d^2/4$
(Figure \ref{fig:schematic}).
The disks interact through elastic collisions only and have ballistic
trajectories in between.
As the disks all have identical mass, this quantity plays no role and is set to
$m=1$.
In this work, we are interested in properties of the hard disk system dependent
on the confluency (the proportion of surface $S$ occupied by the particles)
\begin{equation}
c(t) = n(t) \pi d^2/4 ~,
\end{equation}
where $n(t)$ is the number of particles per unit area at time $t$.
Growth of the population is envisioned as the arrival of randomly spaced
particles according to a predetermined growth curve which -- in keeping with the
model of a cell population in culture -- we take to be the logistic curve 
\begin{equation}\label{eq:growthLogistic}
    n(t) = \frac{k n_0 e^{\rho t}}{k + n_0 \qty( e^{\rho t} - 1 )} ~,
\end{equation}
with $n_0$ and $k$ respectively the initial and the maximal particle densities
and $\rho$ the rate of growth.
We limit ourselves to a confluency $c(t) \ll 1$, effectively considering a
dilute regime for the collisional dynamics. This limitation is also necessary to
make sure that new particles can be inserted easily in the system.

\subsection{Diffusion coefficient for dilute hard disks}

To estimate the motility at a given particle density $n$ we compute the velocity
autocorrelation function (VACF) of a single test particle:
\begin{equation}
    \langle \vb{v}(t) \cdot \vb{v}(t+\tau) \rangle = \langle \cos \theta(\tau) \, s(t)s(t+\tau) \rangle ~,
\end{equation}
where $\vb{v}$ is the 2D velocity of a particle, the brackets indicate ensemble
averaging, $s$ is the norm of the particle's velocity vector (the particle
speed), $\theta(\tau)$ is the angle between its velocities at $t$ and $t+\tau$,
and $\cdot$ is the scalar product.
If the particle moves undisturbed in the short time $\tau$, then $\theta(\tau) =
0$, $s(t+\tau) = s(t)$, and thus $\vb{v}(t)\cdot \vb{v}(t+\tau) = s(t)^2$. If on
the other hand the particle undergoes a collision in this time, the
autocorrelation will depend on the form of the interaction. Assuming isotropic
interactions, there will be a class of systems for which $\theta(\tau)$ can be
modelled as a random uniformly distributed angle independent of the particle
speed, such that the many particle average $\ev{\cos \theta(\tau)}$ and thus
$\ev{\vb{v}(t)\vb{v}(t+\tau)}_{\text{coll}}$ vanishes. Then for a large ensemble
only those particles which have not yet collided will contribute to the VACF.
For a system in thermal equilibrium the occurrence of collisions may reasonably
be modelled as a memoryless stochastic process, so that the inter-collision time
is given by the exponential distribution: $\mathbb{P}\qty{\text{no collision in
}\tau} = e^{-\lambda \tau}$ with $1/\lambda$ the average inter-collision time.
The VACF is thus
\begin{equation}
\label{vacf}
\ev{\vb{v}(t)\cdot \vb{v}(t+\tau) } = \ev{s(t)^2} e^{-\lambda \abs{\tau}} ~.
\end{equation}
%!
By integrating Eq.~\eqref{vacf}, we obtain through the Green-Kubo relation (see
for instance the Ornstein-Uhlenbeck example in Ref.~\cite{gardiner_2004}) the
spatial diffusion coefficient
\begin{equation}
\label{gk}
    \mathcal{D} = \frac{1}{2}\int_0^{\infty} \! \ev{\vb{v}(t)\cdot \vb{v}(t+\tau) } \dd{\tau} = \frac{\ev{s(t)^2}}{2\lambda}
\end{equation}
(where the factor $1/2$ comes from the fact that we work in two spatial
dimensions),
%!
so that in absence of growth we expect a linear mean squared displacement (MSD)
in equilibrium: $\ev{\vb{x}(t)^2} = 4 \mathcal{D} t$.

\subsection{Effective Langevin dynamics}

We establish here the link between the diffusion coefficient computed in the
previous section and the Langevin equation (LE) in two dimensions:
\begin{equation}\label{eq:StandardLangevin}
    \dv{\vb{v}(t)}{t} = -\gamma \vb{v}(t) + \sqrt{2 \eta} \, \xi(t) ~,
\end{equation}
where $\xi(t)$ is white noise~\cite{gardiner_2004}:
\begin{equation}
 \ev{\xi(t)} = 0 \quad \textrm{and} \quad \ev{\xi_i(t)\xi_j(t')} = 
 \delta(t-t')\delta_{i,j} ~,
\end{equation}
$\gamma$ is the friction coefficient and $\eta$ is the noise intensity.
Integration of Eq.~\eqref{eq:StandardLangevin} provides the autocorrelation
function
\begin{equation}\label{eq:velAutocorrBrown}
    \ev{\vb{v}(t)\cdot \vb{v}(t+\tau)} = \frac{2\eta}{\gamma}e^{-\gamma \abs{\tau}} ~.
\end{equation}
For $\tau=0$ we find the second moment of the speed 
\begin{equation}\label{eq:speedMoment2}
    \ev{s(t)^2} = 2\eta/\gamma ~.
\end{equation}
Inserting this back into Eq.~\eqref{eq:velAutocorrBrown} and comparing with
Eq.~\eqref{vacf}, we see that the autocorrelation functions for the Brownian
particle and the ensemble of interacting particles are in fact equal upon
identification of $\lambda = \gamma$.
We use this correspondence to interpret the Langevin dynamics of
Eq.~\eqref{eq:StandardLangevin} as an averaged description of the particle paths
in the interacting ensemble, with the expected time between collisions encoded
in the Brownian friction coefficient as $1/\gamma$.
For the simple dynamical model of hard disks, we can now estimate the friction
and noise intensity directly from the physical parameters so that -- in
equilibrium -- we can use Eq.~\eqref{eq:StandardLangevin} without fitting any
parameter.
In the dilute regime, we can express the mean inter-collision time as a function
of the mean free path $l$, which in turn can be coupled to the particle density
$n$ (with $\sigma=2d$ the collisional cross section)
\cite{chapman1990mathematical}:
\begin{equation}\label{eq:meanFreePath}
    l = \frac{\ev{s}}{\gamma} = 1/\sqrt{2}\sigma n ~.
\end{equation}
At equilibrium, the speed distribution can be shown to approach the Rayleigh
distribution \cite{romanczuk_active_2012} with mean
\begin{equation}
    \label{eq:speedMoment1}
    \ev{s} = \sqrt{\eta\pi/2\gamma} ~,
\end{equation}
and the fluctuation-dissipation relation dictates the relation between $\gamma$
and $\eta$
\begin{equation}\label{eq:fluctuationDissipation}
k_B T = \eta / \gamma ~.
\end{equation}
Thus combining Eqs.~\eqref{eq:meanFreePath}, \eqref{eq:speedMoment1}, and
~\eqref{eq:fluctuationDissipation}, we may write the coefficients in the
Langevin equation in terms of the particle density and system temperature:
\begin{equation}\label{eq:LECollParams}
	\left\lbrace
	\begin{aligned}
		\gamma & = \sqrt{\pi k_B T} \, \sigma n \\
		\eta & = k_B T \gamma
	\end{aligned}
	\right.
\end{equation}
Using the derivations above and Eqs.~\eqref{gk} and \eqref{eq:speedMoment2}, we
obtain the spatial diffusion coefficient
\begin{equation}\label{eq:diffCoeff}
    \mathcal{D} = \eta/\gamma^2 =  \sqrt{k_B T/\pi} \, (\sigma n)^{-1}~.
\end{equation}

Up to this point we have silently assumed that the particle density of the
system is constant, both in the derivation of the velocity autocorrelation as
well as in the assumption of an equilibrium speed distribution. It is however
worth investigating to what extent Eq.~\eqref{eq:LECollParams} holds if the
particle density varies slowly. For example, we might envision a population of
cells undergoing divisions, where the growth rate is slow enough that many
collisions occur in between mitotic events. If then the equilibrium assumption
remains adequate, the LE in Eq.~\eqref{eq:StandardLangevin} can be used with
time-dependent coefficients $\gamma(t)$ and $\eta(t)$ to obtain statistical
properties of the particles subject to the growth function $n(t)$.

%!
While solving the LE with time-dependent coefficients is difficult, we can also
investigate the validity of an overdamped approximation, which corresponds to
taking $\dv*{\vb{v}(t)}{t} = 0$. This permits a straightforward solution of the
LE, as the corresponding Fokker-Planck equation is simply the diffusion equation
with time dependent diffusion coefficient Eq.~\eqref{eq:diffCoeff}. This can be
solved in the usual manner and results in the MSD
\begin{equation}\label{eq:MSDOverdamped}
    \ev{\vb{x}(t)^2} = 4 \int_0^t \mathcal{D}(t') \dd{t'}~.
\end{equation}
%!

\subsection{Simulations}
To test the validity of this approach we also study the hard disk system with
event-driven simulations. We place the disks in a periodic domain and assign
random initial velocities such that the average kinetic energy per particle is
$k_B T$. Due to the randomness of the initial velocity distribution the system
will generally present a nonzero center-of-mass drift, which we remove before
initiating dynamics.
The simulation then proceeds in an event-driven manner. At the start of each
increment, all particles are moved according to their current velocity. Next,
potential collisions (detected as overlapping surfaces) are identified and
sorted in order of occurrence. Collisions are then performed successively --
with newly occurring overlaps being similarly detected, timed, and added to the
queue -- until all overlaps have been treated.
Population growth is implemented by the introduction of new particles at random
positions in the simulation area and fixed times according to the growth curve
in Eq.~\eqref{eq:growthLogistic}.
Newly birthed particles are added with initial speed $\sqrt{2k_B T}$ to ensure
the system evolves isothermally
\footnote{The simulation code, written with the Julia programming language, is
available at {https://github.com/natevmp/particle-crowding }}.
To compare simulation results of the density-varying system with the Langevin
model~\eqref{eq:LECollParams} we compute stochastic realizations of the equation
with the SOSRI algorithm \cite{rackauckas2020stability} using the Julia package
DifferentialEquations.jl \cite{rackauckas2017differentialequations}.

\section{Model results}
\begin{figure}
    \centering
    \includegraphics[width=0.45\textwidth]{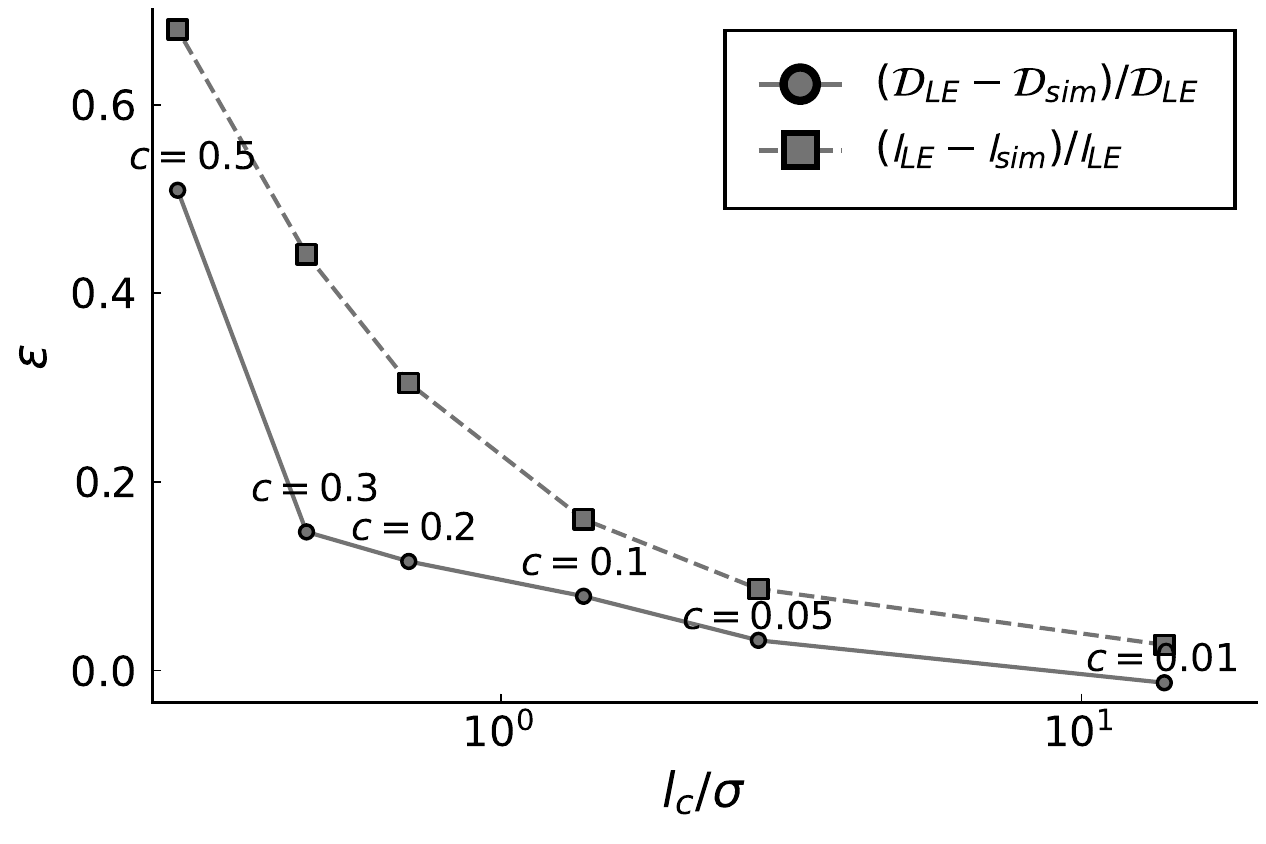}
    \caption{The relative errors on the diffusion coefficient $\mathcal{D}$
        predicted by Eq.~\eqref{eq:diffCoeff} and the mean inter-collision
        distance $l$ predicted by Eq.~\eqref{eq:meanFreePath} with respect to
        simulations at fixed density, as a function of the ratio of the
        inter-collision distance to the collisional cross section. The
        simulations consisted of 2000 individual particles for each confluency
        investigated.}
    \label{fig:fixedDensity}
\end{figure}

For systems where the particle density remains fixed, we find the accuracy of
the Langevin model to depend on the confluency. 
The predicted diffusion coefficient agrees well with simulations of the
hard-disk particle dynamics for $c\leq 0.1$, whereas for higher particle
densities the error grows as the available volume is increasingly occupied
(Figure \ref{fig:fixedDensity}). 
The origin of this error can be found in our assumption of independent
collisions with a well defined mean free path (\ref{eq:meanFreePath}), which
gives a poor representation of the hard-disk system when its value approaches
the order of the collisional cross section.
In the low density limit the model's validity is principally restricted by the
timescale of interest, since if the characteristic collision time $1/\gamma$ is
larger than the timescale the individual particle motion is effectively
ballistic and Brownian motion does not apply. However for a large number of
particles the ensemble statistics remained in agreement even for the lowest
density investigated ($c=0.01$) at a timescale of a hundredth of the
characteristic time.

\begin{figure*}
    \centering
    \includegraphics[width=.67\textwidth]{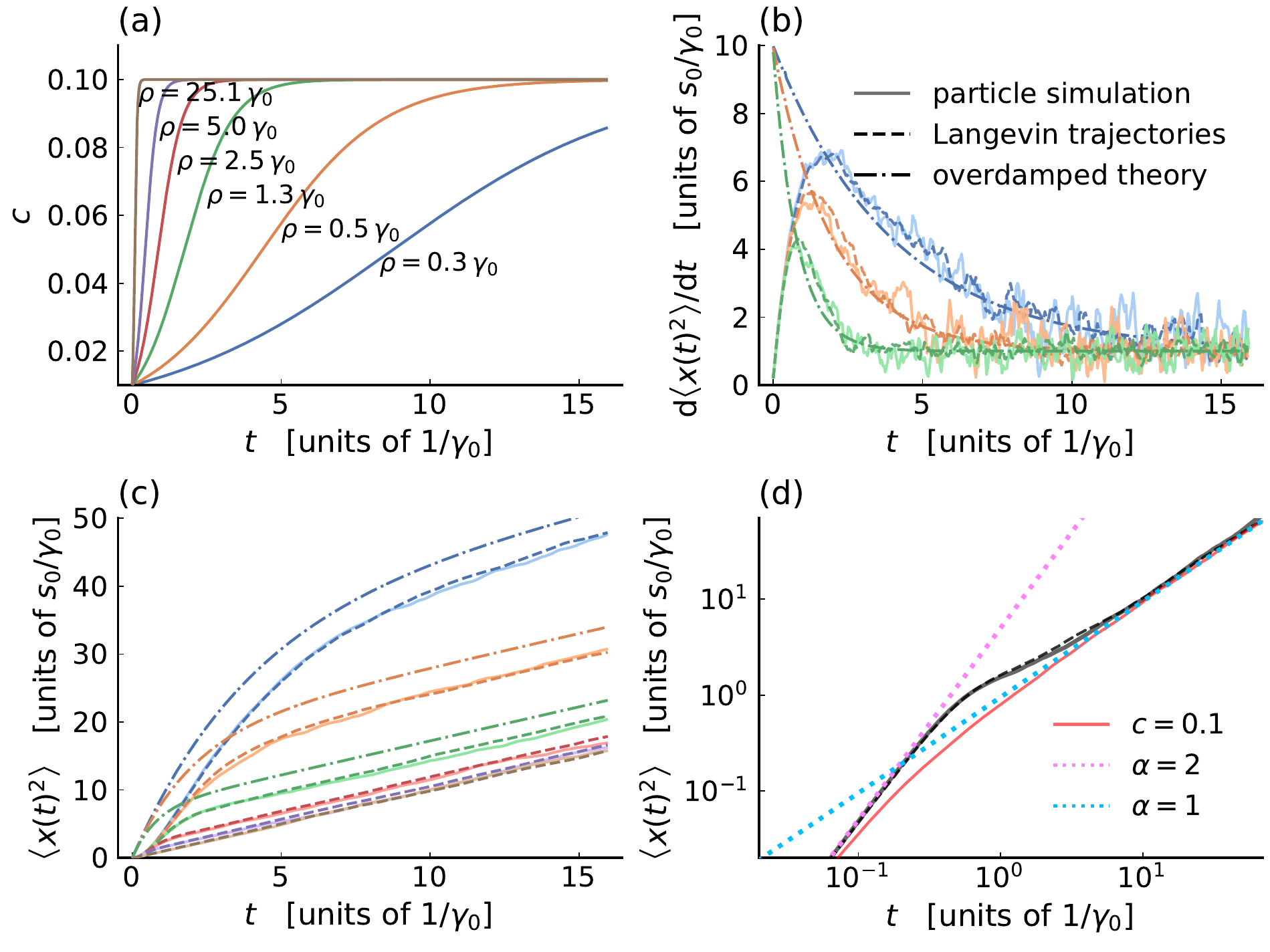}
    \caption{Dynamics of particles in populations with density subject to the
        logistic growth function (\ref{eq:growthLogistic}). Results for
        different growth rates $\rho$ are shown in populations with initial
        confluency $c=0.01$ (1000 particles) and maximal capacity $c=0.1$
        (10'000 particles). Time and distance are presented in units of the
        average inter-collision time for the initial density $1/\gamma_0$ and
        the average inter-collision distance $s_0/\gamma_0$. (a) Confluency over
        time. (b) Derivative of the MSD obtained from particle simulations
        (solid lines), simulations of the LE (dashed lines), and predicted by
        the overdamped approximation (dash-dotted lines), for the slowest growth
        rates used in (a), increasing from top to bottom. (c) MSD under all
        simulated growth rates for particle simulations (same legend as (b); for
        clarity, the overdamped prediction is not shown for the highest growth
        rates). (d) Log-log plot of the MSD under growth rate $\rho =
        5.0\gamma_0$, compared to a simulation at fixed confluency $c=0.1$
        (solid red line). Additional trendlines (dotted) $f_\alpha(t) \propto
        t^\alpha$ are shown to illustrate motion type: $\alpha = 1$ adheres to
        classical diffusion, whereas $\alpha = 2$ implies ballistic motion.}
    \label{fig:msdGrowth}
\end{figure*}

When population growth is introduced the particle density increases and the
Langevin parameters $\gamma$ and $\eta$ are no longer constant in time. We
investigate the validity of the density-dependent model with respect to the
hard-disk system under isothermal growth by simulating the addition of particles
at randomly distributed positions at a rate prescribed by the growth function.
As an illustrative example we consider the case of logistic growth
\eqref{eq:growthLogistic} for different rate parameters $\rho$, shown in Figure
\ref{fig:msdGrowth}a.
%!
We observe that as the density rises the inter-collision time decreases,
resulting in a decreasing variation of the particle MSD during the growth period
and thus apparent subdiffusive motion (Figure \ref{fig:msdGrowth}b-d). Once the
maximal population density is reached the slope of the MSD becomes constant.
Such a linear limit of the functional form of the MSD for large times is an
indicator of classical diffusion (Figure \ref{fig:msdGrowth}c-d).

%!
As intended, simulations of the LE show good agreement with the particle
simulations in the quasi-equilibrium parameter regime (Figure
\ref{fig:msdGrowth}c), where the rate of growth is smaller than the time between
collisions. Furthermore, the overdamped approximation of
Eq.~\eqref{eq:MSDOverdamped} agrees well with the subdiffusive variation of the
MSD, however, as to be expected it does not capture the initial short time
ballistic motion (Figure \ref{fig:msdGrowth}b), resulting in an eventual
overestimation of the displacement (Figure \ref{fig:msdGrowth}c). Interestingly,
the LE model maintains similar accuracy even if the growth rate is significantly
higher, as a growth rate of up to 25 times the inter-collision time was tested.
%!

\section{Discussion}
We have shown how a 2D system of ballistic hard-disk particles subject to
population level density dynamics can be modeled by an effective Brownian
Langevin equation, in which the friction $\gamma$ and force intensity $\eta$ are
made to depend explicitly on the particle density of the system. This
parameter-free dependence is obtained by employing a classic model for the mean
inter-collision distance that -- with the assumption of memoryless collisions --
arises in the velocity autocorrelation function of the LE. By comparison with
simulations of the system, we showed the model to be accurate up to a confluency
of around $0.1$. For higher densities we found the error on the mean
inter-collision distance to grow rapidly with confluency, implying that the LE
could potentially remain accurate under a different model for the free path
lengths.

To test the validity of the LE in the case of a dynamically varying density --
where the assumption of thermal equilibrium is in principle no longer valid --
we investigated a model system where new particles are added according to a
logistic growth function.
The mean-squared displacement under these conditions becomes non-linear,
reflecting the changing dynamic parameters.
Comparing statistics from numerical simulations of the density dependent LE with
the particle simulations showed agreement up to the highest growth rates
investigated.
%!
The good correspondence at such growth rates comes as a surprise, considering
the equilibrium assumptions used to derive the LE. We can think of a number
reasons why this is the case, such as the fact that we insert particles from a
thermal distribution at random locations, thus not promoting spatial
inhomogeneity, and the dilute nature of the gas. However, we have no formal
explanation to elucidate this matter.
%!

While the specific system studied here may appear restrictive, it is
illustrative of the potential for modeling the effect of stochastically
occurring interactions within a density-varying population as an effective
random walk.
%!
A generalization to three spatial dimensions would be straightforward, the main
difference in the derivation being that the Maxwell-Boltzmann distribution must
be taken for the particle speeds. Furthermore, the inclusion of external forces
can be achieved by introducing their relevant potentials in the LE.
The consideration of more complicated interaction effects -- such as for example
aligning forces \cite{vicsek1995novel} or particle-generated flow fields in a
background fluid \cite{baskaran2009statistical} -- likely presents a greater
challenge, as the method described here exploits a simple statistical uniformity
of interactions -- i.e.\ the particle's direction of motion following a
collision modeled as a uniform distribution. Nevertheless, such interactions can
in principle be included through multi-particle potentials in the LE, for which
various analysis procedures exist in the literature (see for example
\cite{tailleur_statistical_2008}, \cite{baskaran2009statistical} or
\cite{stenhammar_continuum_2013}).
%!
Finally, the application of the LE derived here need not be constrained to the
case of ballistically moving particles, as it can simply be added to the LE of
more complex motions, serving as a population effect on the movement of
individual particles. 
%!
With growing interest in biological systems where proliferation is present, this
approach provides a useful alternative to modeling interactions directly.

\end{document}